# Transneptunian Binaries (2018)

Keith S. Noll[1], William M. Grundy[2], David Nesvorný[3], Audrey Thirouin[2]

[1] Goddard Space Flight Center, Greenbelt, MD
[2] Lowell Observatory, Flagstaff, AZ
[3] Southwest Research Institute, Boulder, CO

# 1. Overview

At the time of the Coimbra conference in early 2018, four decades had elapsed since the discovery of the first transneptunian binary companion - Charon, almost two decades since the first discoveries of additional transneptunian binaries (TNBs; e.g. Veillet et al. 2002), and ten years since the last major review of this topic (Noll et al. 2008a). In this chapter we start from the foundation of this earlier work and review new developments in the field that have occurred in the last decade.

By 2008 many of the themes relating to TNBs had already been established and the basic outlines of our current understanding of the transneptunian Solar System and the role of binaries within that population had begun to take shape. Since that time, there has been continued progress in the identification of binaries, determination of mutual orbits, and understanding of system mass, density, rotational state, component colors, and mutual events. Formation models have shifted away from capture scenarios in favor of models that invoke direct gravitational collapse – binaries offer tests of such models while also constraining subsequent dynamical evolution. In 2018, the study of TNBs remains one of the most active areas of progress in understanding the Solar System beyond Neptune.

# 2. Inventory

## 2.1. Direct Imaging

The discovery of new binaries has continued over the last decade, with the majority of new systems being discovered by the *Hubble Space Telescope* (*HST*), as was also true in the previous decade. At the time of the last review, 43 binary systems were known (Noll et al. 2008a); that number has now doubled to 86 transneptunian objects (TNOs) with one or more companions (see compilation[1] maintained by Johnston, 2018).

TNBs have been found in every dynamical population of the Kuiper Belt as shown in Fig. 1. As discussed in more detail below, this includes a large number of binaries in the Cold Classical population (non-resonant, not scattering, low inclination) and significant numbers also in the dynamically hot populations - i.e. objects in mean-motion resonance with Neptune, objects that are scattered by Neptune, detached objects with large perihelia, and high inclination (Hot) Classicals (Elliot et al. 2005, Gladman et al. 2008). Large snapshot surveys with HST have historically been the most productive source of binary discovery accounting for approximately 80% of discoveries. Snapshot observations have been an efficient way to observe a significant number of solar system objects with HST, however, the reduced exposure times dictated by this observing strategy favored brighter objects and observations in a single broad filter. In the last decade, searches with HST have been focused more on follow-up of distinct subpopulations underrepresented in earlier surveys. Overall, HST has observed more than 500 TNOs, a significant fraction of the total transneptunian population with well-known orbits.

---
[1] http://www.johnstonsarchive.net/astro/asteroidmoons.html

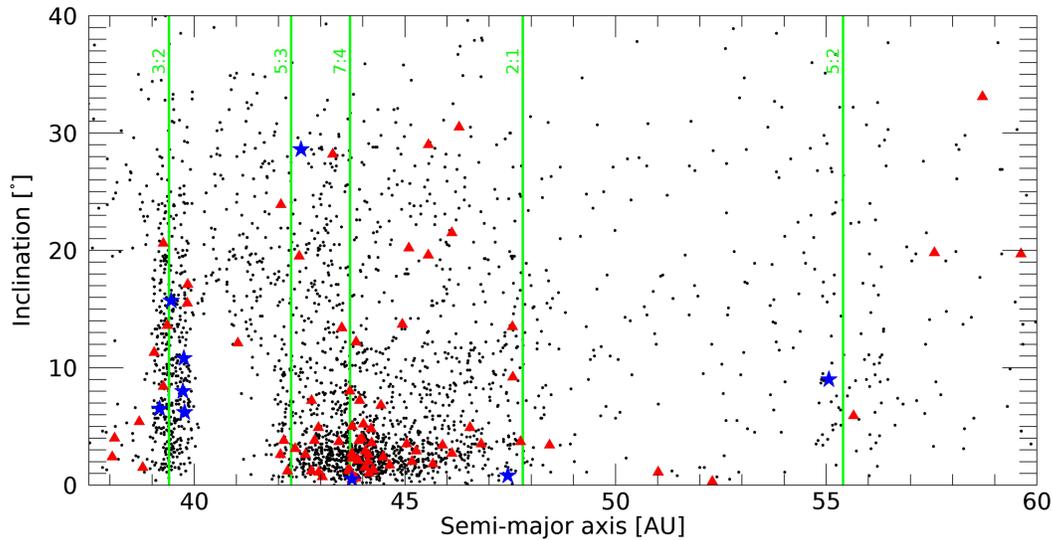

**Fig. 1:** Transneptunian objects with semimajor axes between 38 and 60 AU (black points) are shown in this plot of semimajor axis vs. inclination. Resolved transneptunian binaries are shown as red triangles while possible contact binaries identified from lightcurve analysis are shown as blue stars.

**2.2. Lightcurves**

Lightcurve evidence for elongated bodies, close binaries, and/or bilobed objects (contact binaries), is found in most small body populations including near-Earth objects (Benner et al. 2015), Main Belt Asteroids (Agarwal et al. 2017), Trojans (Ryan et al. 2017), comets (Harmon et al. 2010, Nesvorný et al. 2018), and transneptunian objects (Sheppard & Jewitt 2004, Lacerda et al. 2014; Thirouin & Sheppard 2017, 2018; Thirouin et al. 2014, 2018). For a subset of objects that are large and/or close enough to the Earth or a spacecraft, direct imaging or radar data is available to confirm their bilobed/contact-binary shape. However, for the majority of objects with similar lightcurve signatures, the detailed shape is indeterminate. It is nevertheless clear that such objects are quite common across the Solar System. The transneptunian population is no exception, with early estimates of the fraction of such objects at about 30% (Sheppard & Jewitt 2004, Lacerda 2011).

Potential close/contact binaries in the Kuiper Belt are not directly resolvable due to the small angular separation of the system's components, but candidates can be identified by rotational lightcurves (Thirouin et al. 2017) or by occultations. In the case of the Plutino, 2001 QG$_{298}$, Sheppard & Jewitt (2004) argued that the large 1.14 mag amplitude lightcurve coupled with the 13.77 hr (double-peaked) rotation period was incompatible with an equilibrium fluid body with the required axial ratio. They proposed instead that 2001 QG$_{298}$ is a likely contact binary. The decreased amplitude of 0.7 mag observed in 2010 was compatible with the change in viewing geometry of a contact-binary-like system (Lacerda 2011).

A second example of a TNO with a large lightcurve amplitude (0.85 mag) is 2003 SQ$_{317}$ (Lacerda et al. 2014). This object is of special interest because it is a possible member of the Haumea collisional family. The authors noted the possibility of distinguishing between different hydrostatic equilibrium shape models by monitoring the lightcurve amplitude over multiple years, a technique that can separate elongated ellipsoids and contact binary shapes given the right alignment and sufficient time baseline.

Six additional contact binary candidates have recently been identified from a survey using *Lowell's Discovery Channel Telescope* and the *Magellan Telescope* (Thirouin et al. 2017, Thirouin & Sheppard 2017, Thirouin & Sheppard 2018; (Fig. 2). Preliminary estimates from this survey suggest that as many as ~40-50% of Plutinos but only ~10-25% of Cold Classicals are candidate contact binaries (Thirouin & Sheppard, 2019). This difference, if confirmed with larger numbers of objects, may be an important clue to the dynamical evolution of these populations.

Finally, although it is routine in the literature to refer to the objects we have described as candidate contact binaries, it is worth pointing out that lightcurves are insensitive to concavities and that even rubble piles have sufficient strength that they cannot be treated as hydrostatic fluid bodies (Harris & Werner, 2018). On the other hand, the contact binary shape of the Cold Classical (486958) 2014 MU$_{69}$, first revealed by stellar occultation (Buie et al., 2019) and dramatically confirmed by the New Horizons flyby (Porter et al., 2019) stands as an existence proof and strong validation of both indirect shape determination methods and the models that predict a plentitude of such binaries. The lack of a detectable lightcurve for this object (Benecchi et al. 2019) is a reminder that the lightcurve technique is sensitive to the pole orientation and that many other similar hidden binaries may exist. Continued investigation of the transneptunian contact binary population promises to be a productive observational and theoretical enterprise.

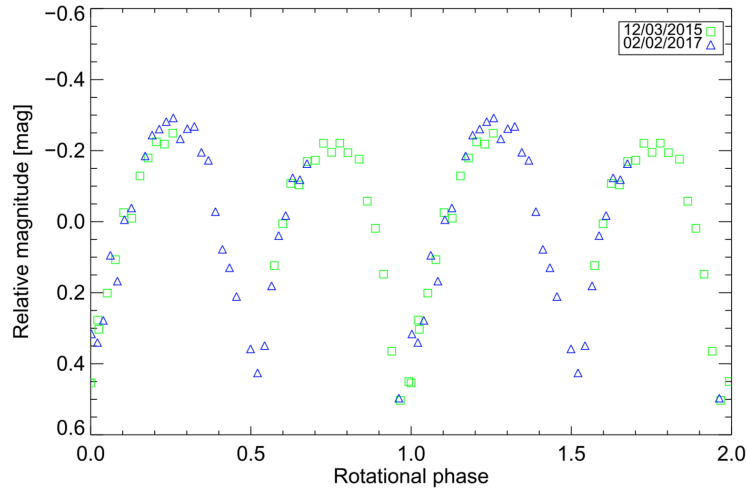

**Fig. 2:** The lightcurve of 2004 TT$_{357}$ shows a large amplitude and a narrow V-shaped minimum that is characteristic of highly elongated bodies including bilobed configurations. The rotational period of 7.79 hr rules out a hydrostatic fluid shape as a possible source of this elongation, leaving a "contact" binary as a possibility (Thirouin et al. 2017).

## 3. Binary Frequency

One of the most frequently asked and seemingly simple questions about TNBs is what fraction of the population they represent. However, any such question must be qualified by knowledge of observational limitations and biases, possible differences between populations, possible size dependencies, knowledge of the separation distribution and regions of orbital stability as well other confounding factors. Any attempt to simply divide the number of detected companions by the number of objects observed will yield a result that is nearly devoid of significance. Below we explore some of the different aspects of determining binary frequency.

While the first TNBs were found from ground-based telescopic observations, the vast majority of known binaries have been discovered by *HST* (Noll et al. 2008a). Only a small fraction of known binaries are ever widely enough separated that they can be observed from ground-based telescopes without adaptive optics (AO), even in the best of seeing. And AO systems must use a laser guide star with stellar appulse because the targets are too faint for wavefront correction. HST detections have heterogeneous instrumental resolution and observational S/N that must be considered when attempting to estimate the frequency of binaries.

In addition to instrumental considerations, the heterogeneity of the sample itself must be considered. The angular size of the Hill sphere is independent of distance for objects with equal mass. But the wide range in albedo (Stansberry et al. 2008) and differing heliocentric eccentricities result in Hill radii that vary depending on an object's density, albedo and orbit. The largest TNOs may retain volatile ices on their surfaces that can cause unusually high albedos. Among smaller TNOs, higher albedos are found among the Cold Classicals than among other dynamical classes (Brucker et al. 2009; Vilenius et al. 2012, 2014). Thus, magnitude-limited surveys will, in general, have a biased sensitivity to secondaries when comparing different classes of objects. We consider two different examples that illustrate issues in appropriately determining binary frequency.

### 3.1. Binary Fraction in the Cold Classicals

One of the earliest published results on the binary fraction in the Kuiper Belt populations was the observation that the fraction of binaries in the Cold Classical population appeared to be greater than that of the dynamically excited populations and, more specifically, the so-called Hot Classicals (Stephens & Noll 2006, Noll et al. 2008b) which overlap the Cold Classicals in semimajor axis space, but have a higher inclination (Levison & Stern 2001, Brown 2001). This observation is one of a group of apparently unique features of the Cold Classicals that, along with their color (Tegler & Romanishin 2000) and the presence of wide binaries (Parker & Kavelaars, 2010), has been used to argue that the Cold Classicals are a relatively undisturbed remnant of the primordial planetesimal disk.

But the high binary fraction in the Cold Classicals is subject to a surprising complication that must be considered carefully. Figure 3 plots 110 Cold Classicals ($i \leq 5.5°$) observed with HST in two different ways. When binned by the absolute magnitude of the unresolved system (i.e. the magnitude reported by ground-based discovery observations), there is a large excess of Cold Classical binaries for 6.15 > H > 4.95 mag (Noll et al. 2014). The Hot Classicals do not show a similar behavior. This effect can be understood as being due to the extreme steepness of the magnitude-frequency distribution (MFD) of the Cold Classicals (Fraser et al. 2014) resulting in the brightest objects being dominated by binaries. Because the observed targets are heavily biased to the brightest available members of the population, the apparent binary fraction can be magnified. However, even at fainter system magnitudes, H > 6.15 mag, the binary frequency among Cold Classicals is ~20% and exceeds that found in the Hot Classicals in the same magnitude range.

The higher albedo for Cold Classicals means that the mass of a Cold Classical is likely to be lower than a Hot Classical of the same absolute magnitude. Because the Hill radius scales with the system mass, the likelihood of finding a secondary at the same separation in terms of Hill radius is, therefore, lower for a Cold Classical than for a Hot Classical of the same absolute magnitude. Thus, the high fraction of Cold Classical binaries appears to be a robust feature of the population.

When the same data are plotted as a function of the absolute magnitude of the primary there is no excess among the largest objects and the frequency is roughly constant down to H~8 mag. Crudely debiasing by scaling all binaries to the faintest absolute magnitude bin and removing undetectable secondaries does not significantly alter this result. Although the dip from H~6.5-7.0 appears significant, Nesvorný et al. (2011) used the lack of a sustained drop off at even fainter magnitudes to constrain collisional grinding among Cold Classicals. Further observations of Cold Classicals, especially fainter ones, are needed to better understand this structure.

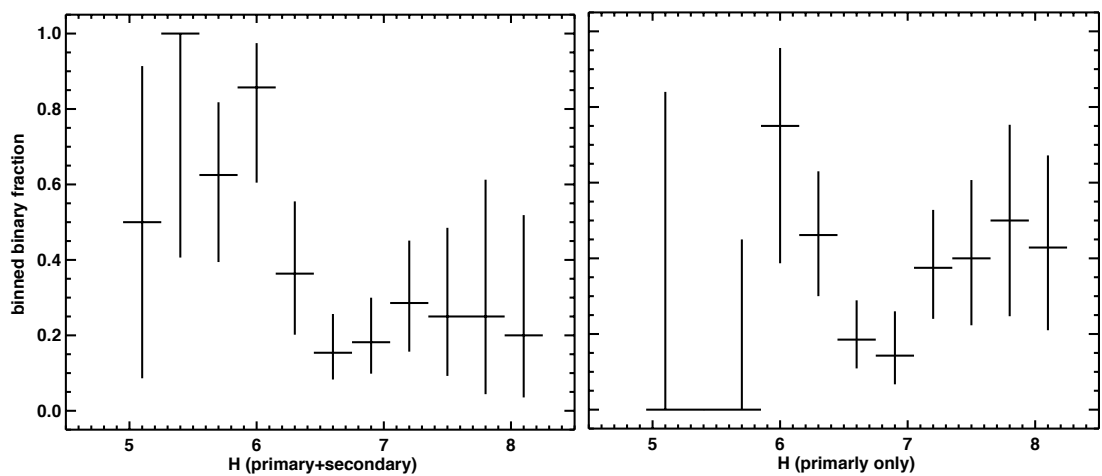

**Fig. 3:** The binned binary fraction of Cold Classical KBOs observed with HST is plotted as a function of the overall system magnitude (left) and by the brightness of the largest component (right). The large excess of binaries among the brightest systems (left) is the product of an extremely steep magnitude-frequency distribution (MFD) that causes the brightest Cold Classicals to be dominated by near-equal mass binaries (Noll et al. 2014). When plotted by the magnitude of the largest component (right), the lack of a drop-off at faint magnitudes is evidence that the observed slope change in the MFD is not due to collisional grinding (Nesvorný et al. 2011)

### 3.2. Resonant Binaries

Binaries found in dynamically excited populations, including Hot Classicals, Scattered Disk, and Resonant populations, are potentially diagnostic indicators of the dynamical perturbations experienced by these objects over their histories. Noll et al. (2012) examined binaries in populations in mean motion resonance with Neptune where a total of 76 Plutinos and 96 objects in other resonances had been observed with HST. As shown in Figure 4, the lack of binary Plutinos at low inclination is particularly striking, with a binary fraction of only $5^{+6}_{-2}$% for $i<12°$ and $5 \leq H \leq 8$ mag. In this same data set, there are no binary Plutinos with $i<5.5°$. This can be compared to $27^{+16}_{-9}$% found for the 2:1 resonance with $i<12°$ in the same absolute magnitude range, a value similar to that found for the Cold Classicals ($29^{+7}_{-6}$%). This result appears to be consistent with the predicted behavior of low-$e$ migration of Neptune and coherent capture into resonances (Murray-Hill and Schlichting 2011) if the 2:1 resonance swept over the Cold Classicals, but the 3:2 resonance did not.

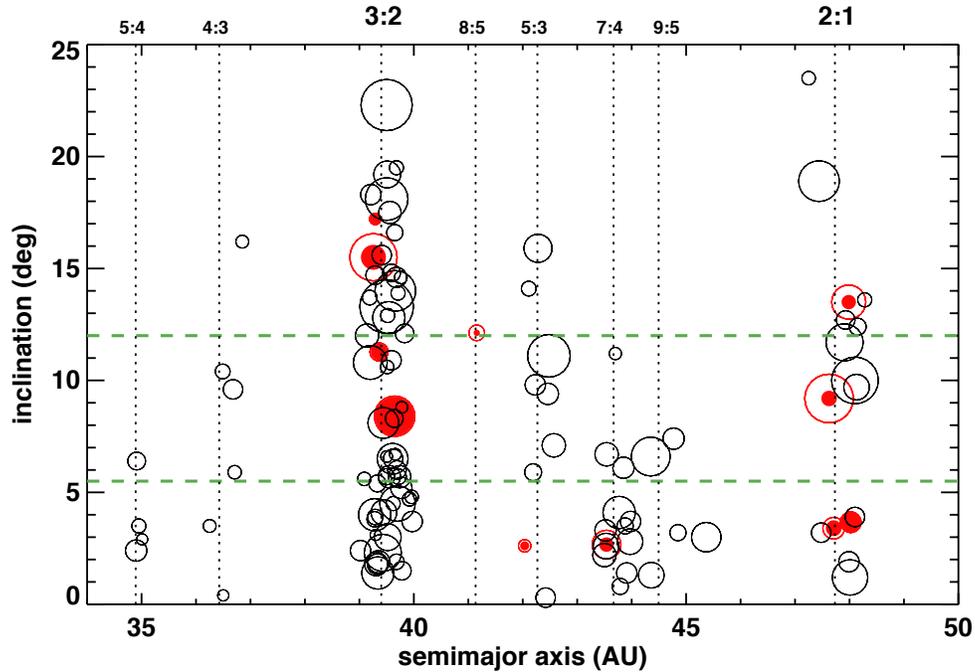

**Fig. 4:** Resonant transneptunians with absolute magnitudes $5 \leq H \leq 8$ mag observed by HST as of 2012 are shown by circular symbols. The size of the symbol scales inversely with absolute magnitude. Objects identified as binaries are shown in red with the primary represented by an open circle and the secondary shown as a filled circle. Resonances with identified objects are shown along the top of the figure. The green dashed horizontal lines at $i=5.5°$ and $i=12°$ mark two proposed "crossover" boundaries of the Cold and Hot Classical inclination distributions. Objects captured by low-$e$ resonance sweeping should have inclinations mostly below these bounds, as described in the text.

## 4. Mutual Orbits

Orbits of transneptunian binaries have mostly been obtained from multi-epoch imaging with sufficiently high spatial resolution to separate the two objects. The widest binaries are separated by a few arcsec, and can be resolved even in natural seeing, but most are separated by much less, requiring adaptive optics or the HST.

Resolved observations at four epochs can, in principle, determine the seven parameters required to fully describe a Keplerian binary orbit (total mass and six orbital elements), but in practice this is a function of how well the observations sample the orbit. The sampling can be improved by using the accumulating collection of data to optimize the timing of each subsequent observation (e.g., Grundy et al. 2008). The mirror ambiguity of an orbit projected onto the sky-plane can eventually be resolved through changing observation geometry, but it takes multiple years for sufficient parallax to accumulate. In some cases, the similarity in brightness between the two components makes it impossible to tell which of them is which from one epoch to the next - more observations are then required to yield a unique solution.

Despite these complications, the last decade has seen significant progress. Of the 86 transneptunian binaries/multiples known, the orbit is fully known for 37. For 11 more, the period, semimajor axis, and eccentricity are known, but the sky plane mirror ambiguity remains to be broken. These orbital parameters are collected in Table 1.

**Table 1: Mutual Orbital Parameters of Transneptunian Binaries**

| Object | $P$ (days) | $a$ (km) | $e$ | Incl.[a] (deg) | $M_{sys}$ ($10^{18}$ kg) | $a/r_H$[b] | ref |
|---|---|---|---|---|---|---|---|
| 26308 1998 SM$_{165}$ | 130.154 | 11374 | 0.4732 | 75.46 | 6.88 | 0.02425 | 1,2 |
| 42355 Typhon - Echidna | 18.9815 | 1580 | 0.507 | 54 | 0.87 | 0.0115 | 3,4 |
| 50000 Quaoar - Weywot | 12.431 | 13330 | 0.021 | - | 1210 | 0.00364 | 5,6 |
| 55637 2002 UX$_{25}$ | 8.3095 | 4750 | 0.18 | 65.0 | 123 | 0.00314 | 6,7 |
| 58534 Logos – Zoe | 309.9 | 8220 | 0.546 | 74.2 | 0.458 | 0.0326 | 1,2,8 |
| 60458 2000 CM$_{114}$ | 24.825 | 2500 | 0.03 | 56 | 2.00 | 0.00667 | 9 |
| 65489 Ceto – Phorcys | 9.560 | 1850 | 0.008 | - | 5.5 | 0.0071 | 10 |
| 66652 Borasisi – Pabu | 46.289 | 4530 | 0.470 | 49.4 | 3.44 | 0.00916 | 11 |
| 79360 Sila - Nunam | 12.51006 | 2770 | 0.026 | 123.3 | 10.8 | 0.00351 | 12 |
| 88611 Teharonhiawako - Sawiskera | 828.8 | 27600 | 0.249 | 127.6 | 2.44 | 0.0581 | 2,13 |
| 90482 Orcus - Vanth | 9.53915 | 9000 | 0.0009 | 106.7 | 635 | 0.004311 | 6,9,14 |
| 119979 2002 WC$_{19}$ | 8.403 | 4090 | 0.20 | - | 77 | 0.00329 | 9 |
| 120347 Salacia - Actaea | 5.49388 | 5720 | 0.010 | 41.4 | 492 | 0.00233 | 4,9 |
| 123509 2000 WK$_{183}$ | 30.913 | 2370 | 0.014 | - | 1.10 | 0.00655 | 2 |
| 134860 2000 OJ$_{67}$ | 22.0585 | 2270 | 0.012 | - | 1.90 | 0.00523 | 15 |
| 136199 Eris - Dysnomia | 15.78590 | 37270 | 0.0062 | 78.5 | 16470 | 0.004724 | 2,16 |
| 148780 Altjira - | 139.56 | 9900 | 0.344 | 25.4 | 3.95 | 0.0182 | 2 |
| 160091 2000 OL$_{67}$ | 347.1 | 7800 | 0.24 | - | 0.32 | 0.035 | 9 |
| 160256 2002 PD$_{149}$ | 1675 | 26800 | 0.588 | 21.9 | 0.54 | 0.099 | 9 |
| 174567 Varda | 5.7506 | 4810 | 0.022 | - | 266 | 0.00232 | 17 |
| 229762 G!kún\|\|'hòmdímà | 11.3147 | 6040 | 0.024 | 32.3 | 136 | 0.00383 | 18 |
| 275809 2001 QY$_{297}$ | 138.118 | 9960 | 0.418 | 161.0 | 4.10 | 0.01855 | 2,9 |
| 341520 Mors-Somnus | 972.2 | 20990 | 0.1494 | 24.3 | 0.776 | 0.0955 | 19 |
| 364171 2996 JZ$_{81}$ | 1500 | 33000 | 0.85 | 11 | 1.3 | 0.090 | 20 |
| 385446 Manwë | 110.18 | 6670 | 0.563 | 49.1 | 1.94 | 0.0167 | 21 |
| 469514 2003 QA$_{91}$ | 10.1089 | 1590 | 0.02 | - | 3.1 | 0.00318 | 9 |
| 469705 ǂKá̦gára | 128.11 | 7700 | 0.69 | 11.2 | 2.2 | 0.0178 | 9 |
| 508788 2000 CQ$_{114}$ | 220.48 | 6940 | 0.095 | 44.4 | 0.545 | 0.0253 | 9 |
| 508869 2002 VT$_{130}$ | 30.761 | 3030 | 0.019 | - | 2.3 | 0.0067 | 9 |
| 1998 WW$_{31}$ | 587.3 | 22620 | 0.819 | 51.7 | 2.66 | 0.04839 | 9,22 |
| 1999 OJ$_{4}$ | 84.115 | 3310 | 0.368 | 57.4 | 0.405 | 0.01448 | 9,15 |
| 1999 RT$_{214}$ | 126.50 | 3400 | 0.30 | 23 | 0.19 | 0.0179 | 9 |
| 2000 CF$_{105}$ | 3900 | 34300 | 0.33 | - | 0.22 | 0.166 | 20 |
| 2000 QL$_{251}$ | 56.449 | 4990 | 0.489 | 134.1 | 3.09 | 0.01097 | 9,15 |
| 2001 QC$_{298}$ | 19.2287 | 3810 | 0.334 | 73.7 | 11.9 | 0.00502 | 1,2 |
| 2001 QW$_{322}$ | 6280 | 102100 | 0.464 | 152.8 | 2.14 | 0.223 | 20 |
| 2001 XR$_{254}$ | 125.58 | 9310 | 0.556 | 21.1 | 4.06 | 0.01687 | 9,15 |
| 2002 XH$_{91}$ | 371.1 | 22400 | 0.71 | 39 | 6.5 | 0.0359 | 9 |
| 2003 QY$_{90}$ | 309.6 | 8550 | 0.663 | 51.4 | 0.52 | 0.0320 | 2 |
| 2003 TJ$_{58}$ | 137.68 | 3830 | 0.516 | 63 | 0.236 | 0.0187 | 2,9 |
| 2003 UN$_{284}$ | 3180 | 54000 | 0.38 | 23.0 | 1.27 | 0.145 | 20 |
| 2004 PB$_{108}$ | 97.020 | 10400 | 0.438 | 83.2 | 9.5 | 0.0148 | 9,15 |
| 2005 EO$_{304}$ | 3600 | 69000 | 0.21 | - | 2.0 | 0.155 | 20 |
| 2006 BR$_{284}$ | 1500 | 25400 | 0.275 | 54.1 | 0.576 | 0.0884 | 20 |
| 2006 CH$_{69}$ | 1420 | 27000 | 0.896 | 133 | 0.8 | 0.081 | 20 |

Table notes:

a. Inclination between the mutual orbit and the heliocentric orbit, where known.

b. Semimajor axis in units of Hill radii, a measure of how tightly bound the orbit is.

c. The first uncertain decimal in each figure is either underlined or the last decimal shown.

[1] Margot et al. 2004, [2] Grundy et al. 2011, [3] Grundy et al. 2008, [4] Stansberry et al. 2012, [5] Fraser et al. 2013, [6] Brown and Butler 2017, [7] Brown 2013, [8] Noll et al 2004a, [9] Grundy et al. 2019, [10] Grundy et al. 2007, [11] Noll et al. 2004b, [12] Grundy et al. 2012, [13] Osip et al. 2003, [14] Brown et al. 2010, [15] Grundy et al. 2009, [16] Brown and Schaller 2007, [17] Grundy et al. 2015, [18] Grundy et al. 2019b, [19] Sheppard et al. 2012, [20] Parker et al. 2011, [21] Grundy et al. 2014, [22] Veillet et al. 2002

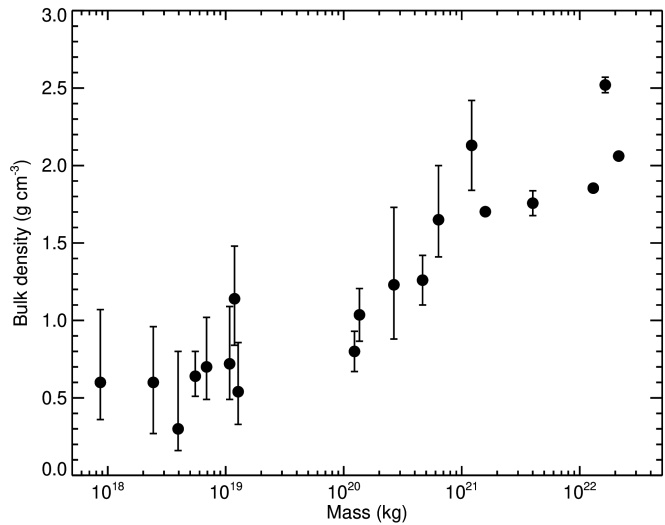

**Fig. 5:** Densities of TNOs versus mass show a smooth increasing trend over the $10^{19}$-$10^{22}$ kg mass range. (Triton and Charon are included because they fall in this mass range and have well determined densities).

## 5. Properties derived from orbits

### 5.1. Mass and Density

The total mass of a binary system can be calculated once the period and semi-major axis of the binary orbit are known. Mass is a fundamental property, providing valuable constraints on dynamics, as well as composition and internal structure, via the bulk density computed by dividing the mass by the estimated total volume of the bodies in the system. Object sizes, necessary to derive density, have been challenging to obtain. Observations at thermal infrared wavelengths can constrain radii to about ±10% precision, translating to a ±30% uncertainty in volume and density. As shown in figure 5, TNBs with system masses, $M_{sys}$, of ~$10^{19}$ kg and less, corresponding roughly to component diameters of d< 300 km, have very low densities, mostly less than 1000 kg m$^{-3}$. More massive objects, $M_{sys}$ >$10^{21}$ kg, d>900 km, have higher densities, generally greater than 1500 kg m$^{-3}$. In between, diverse densities are seen among the mid-sized bodies. Brown (2013) argued that the objects in the 300-900 km diameter size range are too large to have retained much internal porosity, and suggested that their lower densities result from predominately ice-rich compositions, with more rock-rich compositions in the larger bodies accounting for their higher densities. However, a plot of density versus mass (Fig. 5) shows a consistent trend suggesting that a gradual reduction in pore space with increasing mass can explain the variation in density without resorting to *ad hoc* assumptions about composition. Looking ahead, stellar occultations enabled by the GAIA astrometric catalog have the potential to provide much more accurate sizes, and consequently better densities that will allow more detailed study of the densities of TNBs.

**5.2. Orbital Eccentricity, Inclination, and Separation**
The orbital properties of binaries can provide information about their formation circumstances as well as their subsequent dynamical evolution. For instance, circular orbits occur mostly among tightly bound binaries, (Fig. 6), which is not a surprise, since these are the systems most prone to tidal evolution. Wide binaries with appreciable inclinations between their mutual orbits and heliocentric orbits are subject to the Kozai-Lidov effect (Kozai 1962; Lidov 1962) where the gravity of the distant Sun perturbs the mutual orbit, causing an oscillation between $e$ and $i$. Grundy et al. (2011) argued that the broad peak of eccentricities around $e = 0.5$ was consistent with a population undergoing these oscillations. Because wide binaries are easier to discover and characterize than tight ones, it is likely that the tight circular orbits are underrepresented among the known orbits relative to the wider, more eccentric ones.

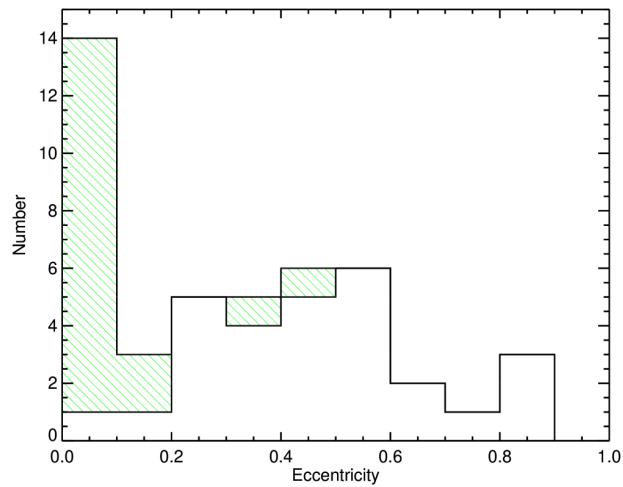

**Fig. 6:** Distribution of eccentricities of known binary orbits. Close binaries, *i.e.* those with $a/r_H < 0.1$, are indicated by green hatching. Only one of the nearly-circular orbits ($e < 0.1$) is not a close binary. At higher eccentricities most binaries are wider, although four close binaries also have eccentricities higher than 0.1.

The observation that there are considerably more prograde orbits than retrograde ones (Fig. 7; Grundy et al. 2011, 2019) is contrary to expectation from various capture scenarios (e.g., Goldreich et al. 2002; Schlichting and Sari 2008; Kominami et al. 2011) that favor retrograde orbits, or, at most, equal proportions of the two. Collisions should also not produce an excess of prograde systems. Gravitational collapse from locally concentrated regions of the nebula, however, might preferentially produce prograde binaries (Nesvorný et al. 2010, Grundy et al. 2019).

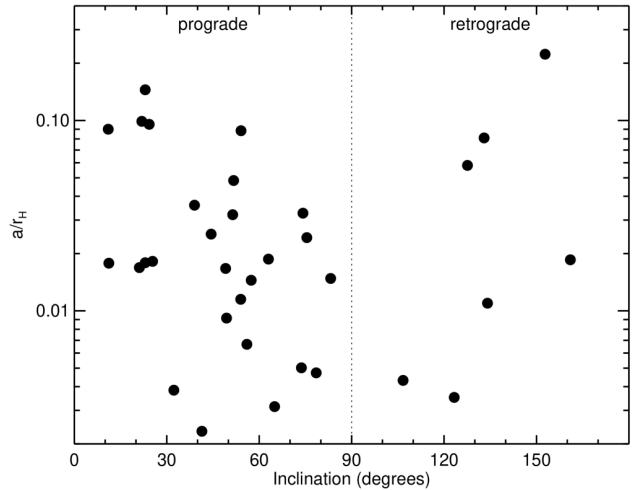

**Fig. 7:** Prograde orbits vastly outnumber retrograde orbits, except among the widest binaries. There is also a shortage of inclinations near 90° among the wider binaries.

The widest binaries, those with $a/r_H$ greater than about 0.05, do not show a strong prograde-retrograde asymmetry but do lack orbits with inclinations near 90°. These highly inclined orbits are the ones that undergo the largest amplitude Kozai oscillations, leading to times of especially high eccentricity that could accelerate tidal evolution (e.g., Perets and Naoz 2009; Naoz et al. 2010; Porter and Grundy 2012), converting them into much tighter binaries, or perhaps leading to their becoming unbound.

A plot of $a/r_H$ versus the excitation of the heliocentric orbit shows that most of the TNBs on more perturbed orbits are tightly bound (Fig. 8). This may be because the process of perturbing them into their present-day heliocentric orbits also dissociated most of the less tightly bound binaries. An exception is Mors-Somnus, a 3:2 resonant object (Sheppard et al. 2012) that is also a wide binary. Its existence suggests that non-destructive paths to excited heliocentric orbits, such as resonance sweeping, may have affected TNO populations.

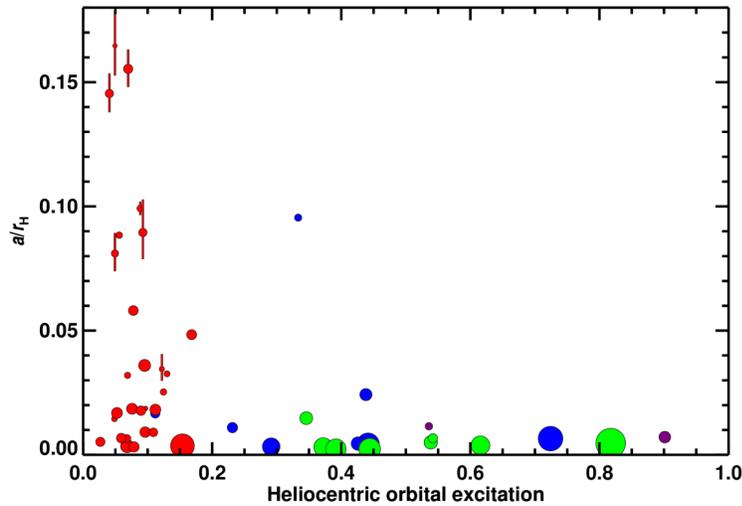

**Fig. 8:** Binary orbital looseness, as measured by $a/r_H$, versus the excitation of the heliocentric obit, as measured by $\sqrt{sin(i_\odot)^2 + e_\odot^2}$, where $i_\odot$ and $e_\odot$ are the inclination and eccentricity of the heliocentric orbit. Colors indicate dynamical class: red for Classical TNOs in low eccentricity, low inclination orbits, blue for objects in mean motion resonance with Neptune, and green for scattered disk objects. The two purple objects are Centaurs, effectively scattered disk objects that have been perturbed into short-lived orbits crossing those of one or more of the giant planets.

### 5.3. Mutual Events and Occultations

Two additional channels for obtaining physical information that have begun to be exploited are mutual events and occultations. Mutual events occur twice in each TNB orbit when the binary orbit plane and the line of sight to Earth are nearly aligned. The duration of the individual mutual events and mutual event season depend on the particulars of the binary orbit, but are on the order of hours for individual events and ~years for the season. With >40 TNB orbits known, it is reasonable to expect that one or two TNBs will have observable mutual events per decade.

Mutual events of the binary (79360) Sila-Nunam were predicted by Grundy et al. (2012). Benecchi et al. (2014) successfully observed one such event in February 2013 from multiple telescope facilities. Grundy et al. (2014) also predicted mutual events for (385446) Manwë-Thorondor in 2015-2017. Attempts to observe Manwë-Thorondor with the SOAR telescope in August 2016 were reported by Rabinowitz et al. (2016), but final analysis remains pending. Fabrycky et al. (2008) and Ragozzine and Brown (2009) described a series of mutual events involving Haumea and one of its satellites, Namaka and attempted to observe one event using HST. Results of this observation were inconclusive. Fully realizing the potential of TNB mutual events will require more significant investments of large telescope observing time than has been available up to now.

Occultations by TNBs are another avenue of study that has only begun to be exploited. Sickafoose et al. (2019) report a ground-based occultation of Vanth, the satellite of Orcus, leading to diameter determination of 443±10 km assuming a spheroidal body. The occultation-derived contact-binary shape of (486958) 2014 MU$_{69}$ (Buie et al. 2019) is another poignant example of the potential of occultation observations. The availability of precise stellar astrometric data from Gaia (e.g. Brown et al. 2018) promises to greatly expand the opportunities for similar observations in the future.

## 6. Colors

One of the most significant results to emerge from the study of TNBs has been the near-identity of component colors (Noll et al. 2008a, Benecchi et al 2009). This identity coupled with the wide range of colors of TNBs puts strong constraints on both formation mechanisms and subsequent color evolution. Capture models are limited to homogeneous regions and/or times within the larger potential color range seen in the current Kuiper Belt. Subsequent physical evolution is also limited; in particular, non-disruptive collisions do not appear to result in color changes which would be seen as rotational color variability and/or variations in component colors.

Exceptions to binary component color identity among the largest TNOs and their satellites are attributable to highly volatile ices that are retained on the larger body, but are not stable on smaller components (e.g. Parker et al. 2016). More generally, it is possible that the satellites of the largest TNOs may have originated from impacts (e.g. Leinhardt et al. 2010), rather than by coaccretion or similar co-formation mechanisms (see section 7) and thus could have more global compositional differences.

Sheppard et al. (2012) noted that the wide Plutino binary, Mors-Somnus, has a very red color and argued for capture from the Cold Classical population based on the combination of this color and the near-equal component sizes. Fraser et al. (2017) used a similar color-based argument to leverage the neutral colors of several wide Cold Classical binaries into a more global conclusion on the compositional segregation of the protoplanetary disk and the ubiquity of binaries. However, the wide range of colors found in all transneptunian populations (e.g. Peixinho et al. 2015) greatly limits the use of color as an identifier capable of linking an individual object to a parent population. A much larger data set will be required in order to test these and similar ideas.

## 7. Formation Scenarios

Transneptunian binaries are an important tracer of planetesimal formation in the outer Solar System and understanding how binaries formed is a critical test of planetesimal formation models. For example, if formed by capture (e.g., Goldreich et al. 2002), they constrain the dynamical conditions in the early protoplanetary disk (Schlichting & Sari 2008). However, as noted above, the matching colors of binary components and prograde orbital orientations argue against this formation model.

An appealing alternative is a model in which planetesimals formed by the streaming instability (SI; Youdin & Goodman 2005). The results of detailed hydrodynamical simulations (Johansen et al. 2009, Simon et al. 2017, Li et al. 2018) show that the SI clumps become gravitationally bound and result in a characteristic planetesimal size of ~100 km (depending on the details of disk parameters; Johansen et al. 2009).

The existing SI simulations also indicate that the collapsing clouds have vigorous rotation and their initial angular momentum typically exceeds, by a large margin, that of a critically rotating Jacobian ellipsoid with density $\rho$=1000 kg m$^{-3}$. This large excess of angular momentum that must be lost can produce a binary (Nesvorný et al. 2010; hereafter NYR10) with most of the angular momentum being deposited into the binary orbit.

NYR10 studied the collapse stage and Figure 8 summarizes some results of the gravitational collapse. Equal-size KBO binaries ($R_2/R_1$>0.5, where $R_1$ and $R_2$ are the radii of primary and secondary components) form in the simulations if the initial clumps contract below the Hill radius ($R \leq 0.6\ R_H$) and/or have fast rotation ($\omega \geq 0.5\ \omega_c$). This is roughly consistent with the results of the SI simulations where the clumps are small $R \sim 0.1\ R_H$ and rotate as fast as they could. The gravitational collapse simulations also predict large separations between binary components (~1000 to ~10$^5$ km) and a broad distribution of binary eccentricities. Depending on the local disk conditions, the gravitational collapse may be capable of producing up to 100% binary fraction (for $\omega \geq 0.5\ \omega_c$). Because the binary components form from the local composition mix, the components should have identical compositions and colors, consistent with observations (see section 6).

Another argument can be based on the inclination distribution of binary orbits (see section 5). The SI simulations can be used to extract the orientation of the angular momentum vectors of individual clumps. NYR10 showed that the initial angular momentum vectors are a good proxy for the final orientation of binary orbits, allowing the SI results to be directly compared with observed binary orbits (Table 1); the results show a strikingly good match (Fig. 9).

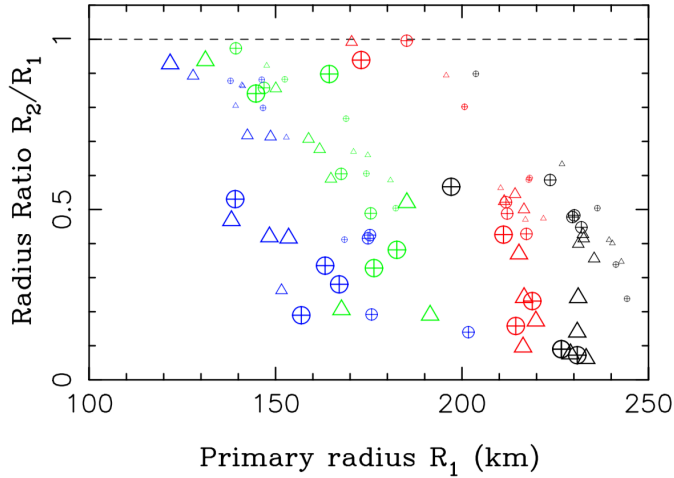

**Fig. 9:** The plot shows the sizes of binary components obtained in 96 simulations of gravitational collapse. Different initial conditions were used in each of these simulations. The initial rotation was assumed to be prograde (triangles) or retrograde (crossed circles) with respect to the orbital motion. The initial size of clumps, $R$, was set to be a fraction of the Hill radius at 30 AU, $R_H$ (0.4, 0.6 and 0.8 $R_H$ as indicated by the symbol size). The initial rotation was assumed to be a fraction of the critical frequency $\Omega_c$ (see NYR10 for definition): 0.1 (black), 0.25 (red), 0.5 (green) and 0.75 $\Omega_c$ (blue). For each setup, we performed four different simulations with slightly altered distributions of $N=10^5$ superparticles.

## 8. Future Observations and Summary

The entire field of study of transneptunian space stands on the cusp of a major revolution in data availability to be brought about by new surveys that will greatly expand the number of known transneptunian objects. Because these surveys will be carried out by ground-based telescopes with limited angular resolution, a proportionally large jump in the number of known binaries will come about as well, but more slowly as the result of painstaking follow ups with advanced AO or space-based telescopes. More binary orbits will also be determined, both from already-known binaries and yet-to-be discovered systems. Taken in whole, on the timescale of the next decade, these developments promise improved statistics that will bring some of the conclusions reached so far into sharper focus.

Perhaps the greatest potential for the study of TNBs lies in the further observations of TNO lightcurves and of stellar occultations. Both of these techniques make it possible to explore close-in binaries – a population that cannot be reached by direct imaging. These efforts should be bolstered by the incontrovertible evidence delivered by the New Horizons mission that a heretofore under-appreciated population of primordial binary objects awaits our study.

Finally, the expansion of our understanding of TNBs opens the door to comparisons with other related population including the Jupiter and Neptune Trojans, Centaurs, and comets. At least one of the targets of the Lucy mission, the Patroclus-Menoetius binary, has properties that look very much like those of many of the known TNBs (e.g. Buie et al. 2015). Answering whether this signals a genetic link or is simply a matter of chance will rely both on spacecraft data and a deep understanding of the binary populations of the outer Solar System.

## Acknowledgments

W.M.G. and K.S.N. gratefully acknowledge support from NASA/ESA Hubble Space Telescope programs 13404, 13668, 13692, and 15233. Support for these programs was provided by the National Aeronautics and Space Administration (NASA) through grants from the Space Telescope Science Institute (STScI), operated by the Association of Universities for Research in Astronomy, Inc., (AURA) under NASA contract NAS 5-26555. Additional support was provided to W.M.G. through NASA Keck PI Data Awards, administered by the NASA Exoplanet Science Institute. D.N.'s work is supported by NASA's Emerging Worlds. A.T. is partly supported by the National Science Foundation (NSF), grant number AST-1734484. A.T. also acknowledges Scott Sheppard for his contribution to the contact binary search and characterization.